\documentclass[prd,aps,nofootinbib,preprint,tightenlines]{revtex4}

\newcommand{\checked}[1]{}

\usepackage{graphicx}

\newcommand{\beq}{\begin{equation}}
\newcommand{\eeq}{\end{equation}}
\newcommand{\bqa}{\begin{eqnarray}}
\newcommand{\eqa}{\end{eqnarray}}
\def\simge{\mathrel{
    \rlap{\raise 0.511ex \hbox{$>$}}{\lower 0.511ex \hbox{$\sim$}}}}
\def\simle{\mathrel{
    \rlap{\raise 0.511ex \hbox{$<$}}{\lower 0.511ex \hbox{$\sim$}}}}

\begin{document}

\title{The heavy-quark potential in an anisotropic plasma}


\author{Adrian Dumitru$^a$, Yun Guo$^{b,c}$ and Michael Strickland$^a$}
\affiliation{$^a$Institut f\"ur Theoretische Physik,
Johann Wolfgang Goethe Universit\"at,
Max-von-Laue-Str.\ 1,
D-60438 Frankfurt am Main, Germany\\
$^b$Helmholtz Research School,
Johann Wolfgang Goethe Universit\"at,
Max-von-Laue-Str.\ 1, D-60438 Frankfurt am Main, Germany\\
$^c$Institute of Particle Physics, Huazhong Normal University,
Wuhan 430079, China
\vspace*{1cm}
}

\begin{abstract}
We determine the hard-loop resummed propagator in an anisotropic QCD
plasma in general covariant gauges and define a potential between
heavy quarks from the Fourier transform of its static limit.
We find that there is stronger attraction on distance scales on the
order of the inverse Debye mass for quark pairs aligned along the
direction of anisotropy than for transverse alignment.
\end{abstract}
\maketitle

\small

\section{Introduction}

Information on quarkonium spectral functions at high temperature has
started to emerge from lattice-QCD simulations; we refer to
ref.~\cite{Jakovac:2006sf} for recent work and for links to earlier
studies. This has motivated a number of attempts to understand the
lattice measurements within non-relativistic potential models
including finite temperature effects such as
screening~\cite{Mocsy:2005qw}. A detailed discussion of the properties
of the heavy-quark potential in the deconfined phase of QCD is given
in ref.~\cite{Mocsy:2007yj}, which also provides a comprehensive list
of earlier work. Also, Laine {\it et al.} have recently derived a
Schr\"odinger equation for the finite-temperature Wilson loop to
leading order within ``hard-thermal loop'' (HTL) resummed perturbation
theory by analytic continuation to real
time~\cite{Laine:2006ns}. Aside from the well-known screened Debye
potential, their result includes an imaginary part due to Landau
damping of low-frequency modes of the gauge field, corresponding to a
finite life-time of quarkonium states.

The present paper is a first attempt to consider the effects due to a
local anisotropy of the plasma in momentum space on the heavy-quark
potential. Such deviations from perfect isotropy are expected for a
real plasma created in high-energy heavy-ion collisions, which
undergoes expansion. The HTL propagator of an anisotropic plasma has
been calculated in time-axial gauge in
ref.~\cite{Romatschke:2003ms}. We derive the result for general
covariant gauges, which allows us to define a non-relativistic
potential via the Fourier transform of the propagator in the static
limit.

\section{Hard-Thermal-Loop self-energy in an anisotropic plasma}
\label{htl-sec}

The retarded gauge-field self-energy in the hard-loop approximation is
given by~\cite{ThomaMrow}
\beq
\Pi^{\mu \nu}(p)= g^2 \int \frac{d^3 {\bf k}}{(2\pi)^3} \,
v^{\mu} \frac{\partial f({\bf k})}{\partial k^\beta}
 \left( g^{\nu \beta} -
\frac{v^{\nu} p^\beta}{p\cdot v + i \epsilon}\right) \; .
\label{selfenergy1}
\eeq
Here, $v^{\mu} \equiv (1,{\bf k}/|{\bf k}|)$ is a light-like vector
describing the propagation of a plasma particle in space-time. The
self-energy is symmetric, $\Pi^{\mu\nu}(p)=\Pi^{\nu\mu}(p)$, and
transverse, $p_\mu\Pi^{\mu\nu}(p)=0$.

In a suitable tensor basis the components of $\Pi^{\mu\nu}$ can be
determined explicitly. For anisotropic systems there are more
independent projectors than for the standard equilibrium
case~\cite{Romatschke:2003ms}. Here, we extend the tensor basis used
in~\cite{Romatschke:2003ms} to a four-tensor basis appropriate for use
in general covariant gauges. Specifically,
\beq \label{eq:A_munu}
A^{\mu \nu}= -g^{\mu\nu}+\frac{p^\mu
p^\nu}{p^2}+\frac{\tilde{m}^\mu \tilde{m}^\nu}{\tilde{m}^2}
\eeq
\beq
B^{\mu \nu}= -\frac{p^2}{(m\cdot p) ^2}\frac{\tilde{m}^\mu
\tilde{m}^\nu}{\tilde{m}^2}
\eeq
\beq
C^{\mu \nu}= \frac{\tilde{m}^2p^2}{\tilde{m}^2p^2+(n\cdot
p)^2}[\tilde{n}^\mu
\tilde{n}^\nu-\frac{\tilde{m}\cdot\tilde{n}}{\tilde{m}^2}(\tilde{m}^\mu
\tilde{n}^\nu+\tilde{m}^\nu
\tilde{n}^\mu)+\frac{(\tilde{m}\cdot\tilde{n})^2}{\tilde{m}^4}\tilde{m}^\mu
\tilde{m}^\nu]
\eeq
\beq   \label{eq:D_munu}
D^{\mu \nu}= \frac{p^2}{m\cdot p}
\left[ 2\frac{\tilde{m}\cdot\tilde{n}}{\tilde{m}^2}\tilde{m}^\mu
\tilde{m}^\nu-
\left(\tilde{n}^\mu \tilde{m}^\nu+\tilde{m}^\mu
\tilde{n}^\nu\right) \right]~.
\eeq
Here, $m^{\mu}$ is the heat-bath vector, which in the local rest frame
is given by $m^{\mu}=(1,0,0,0)$, and
\begin{equation}
\tilde{m}^\mu=m^{\mu}-\frac{m\cdot p}{p^2} \,p^\mu
\end{equation}
is the part that is orthogonal to $p^\mu$.

The direction of anisotropy in momentum space is determined by the
vector
\begin{equation}
n^{\mu}=(0,{\bf n})~,
\end{equation}
where ${\bf n}$ is a three-dimensional unit vector. As before,
$\tilde{n}^\mu$ is the part of $n^\mu$ orthogonal to $p^\mu$.

The self-energy can now be written as
\begin{equation}
\Pi^{\mu\nu}=\alpha A^{\mu\nu}+\beta B^{\mu\nu} + \gamma
C^{\mu\nu} + \delta D^{\mu\nu} ~.
\end{equation}
In order to determine the four structure functions explicitly we
need to specify the phase-space distribution function. We employ the
following {\em ansatz}:
\begin{equation}
f({\bf p}) = f_{\rm iso}\left(\sqrt{{\bf p}^2+\xi({\bf p}\cdot{\bf
n})^2} \right) ~.  \label{eq:f_aniso}
\end{equation}
Thus, $f({\bf p})$ is obtained from an isotropic distribution $f_{\rm
iso}(|\bf{p}|)$ by removing particles with a large momentum component
along $\bf{n}$. The function $f_{\rm iso}(|\bf{p}|)$ should decrease
monotonically with $|\bf{p}|$, so that the square of the Debye mass
defined in eq.~(\ref{eq:Debye}) below is guaranteed to be positive;
however, in the real-time approach employed here, the distribution
$f_{\rm iso}$ need not necessarily be thermal.

The parameter $\xi$ determines the degree of anisotropy, $\xi = (1/2)
\, \langle {\bf p}_\perp^2\rangle / \langle p_z^2\rangle -1$, where
$p_z\equiv \bf{p\cdot n}$ and ${\bf p}_\perp\equiv \bf{p-n (p\cdot
n)}$ denote the particle momentum along and perpendicular to the
direction ${\bf n}$ of anisotropy, respectively. {\em If} $f_{\rm
iso}$ is a thermal ideal-gas distribution {\em and} $\xi$ is small
then $\xi$ is also related to the shear viscosity of the plasma; for
example, for one-dimensional Bjorken expansion~\cite{Asakawa:2006tc}
\beq \label{eq:xi_eta}
\xi = \frac{10}{T\tau} \frac{\eta}{s}~,
\eeq
where $T$ is the temperature, $\tau$ is proper time, and $\eta/s$ is
the ratio of shear viscosity to entropy density. In an expanding
system, non-vanishing viscosity implies finite momentum relaxation
rate and therefore an anisotropy of the particle momenta.

Since the self-energy tensor is symmetric and transverse, not all of
its components are independent. We can therefore restrict our
considerations to the spatial part of $\Pi^{\mu\nu}$,
\begin{equation} \label{eq:Pi_spatial}
\Pi^{i j}(p,\xi) = m_{D}^2 \int \frac{d \Omega}{4 \pi} v^{i}
\frac{v^{l}+\xi({\bf v}\cdot{\bf n}) n^{l}}{%
(1+\xi({\bf v}\cdot{\bf n})^2)^2} \left( \delta^{j l}+\frac{v^{j}
p^{l}}{p\cdot v + i \epsilon}\right) ~,
\end{equation}
and employ the following contractions:
\begin{eqnarray}
p^{i} \Pi^{ij} p^{j} & = & {\bf p}^2 \beta ~, \nonumber \\
A^{il}{n}^{l} \Pi^{ij} p^{j} & = &({\bf p}^2-(n\cdot p)^2)\delta ~,
       \nonumber \\
A^{il}{n}^{l} \Pi^{ij} A^{jk}{n}^{k} & = &
 \frac{{\bf p}^2-(n\cdot p)^2}{{\bf p}^2} (\alpha+\gamma)~ , \nonumber \\
{\rm Tr}\,{\Pi^{ij}} & = & 2\alpha +\beta +\gamma ~.
\label{contractions}
\end{eqnarray}
The Debye mass $m_D$ appearing in eq.~(\ref{eq:Pi_spatial}) is given by
\beq \label{eq:Debye}
m_D^2 = -{g^2\over 2\pi^2} \int_0^\infty d \rho \,
  \rho^2 \, {d f_{\rm iso}(\rho) \over d\rho} ~,
\eeq
where $\rho\equiv |{\bf p}|$. We do not list the rather cumbersome
explicit expressions for the four structure functions $\alpha$,
$\beta$, $\gamma$, and $\delta$ here since they have already been
determined in ref.~\cite{Romatschke:2003ms}.

In principle, the tensor basis~(\ref{eq:A_munu}-\ref{eq:D_munu}) could
be chosen differently, such that the individual tensors have a simpler
structure. For example, one could choose
\beq C^{\mu \nu}= \tilde{n}^\mu
\tilde{n}^\nu-\frac{\tilde{m}\cdot\tilde{n}}{2\tilde{m}^2}(\tilde{m}^\mu
\tilde{n}^\nu+\tilde{m}^\nu \tilde{n}^\mu)\eeq
\beq D^{\mu \nu}=
\frac{(\tilde{m}\cdot\tilde{n})^2}{\tilde{m}^4}\tilde{m}^\mu
\tilde{m}^\nu-\tilde{n}^\mu \tilde{n}^\nu \eeq
However, in the basis~(\ref{eq:A_munu}-\ref{eq:D_munu}) the spatial
components of $\Pi^{\mu\nu}$ are identical to those from
ref.~\cite{Romatschke:2003ms} and so we can avoid the rather tedious
re-evaluation of the four structure functions.

\section{Propagator in covariant gauge in an anisotropic plasma}
\label{prop-sec}

From the above result for the gluon self-energy one can obtain the
propagator $i\Delta^{\mu\nu}_{ab}$. It is diagonal in color
and so color indices will be suppressed.  In covariant gauge, its
inverse is given by
\begin{eqnarray}
&&\left(\Delta^{-1}\right)^{\mu \nu}(p,\xi)= -p^2 g^{\mu \nu} +p^\mu p^\nu
-\Pi^{\mu\nu}(p,\xi)-\frac{1}{\lambda}p^\mu p^\nu \nonumber\\
&&\qquad\qquad\quad =(p^2-\alpha) A^{\mu\nu}+(\omega^2-\beta)
B^{\mu\nu} - \gamma C^{\mu\nu} - \delta
D^{\mu\nu}-\frac{1}{\lambda}p^\mu p^\nu
\end{eqnarray}
where $\omega\equiv p\cdot m$ and $\lambda$ is the gauge
parameter. Upon inversion, the propagator is written as
\begin{equation}  \label{eq:Delta_munu}
\Delta^{\mu \nu}(p,\xi)= \alpha^{\prime} A^{\mu\nu}+\beta^{\prime}
B^{\mu\nu} + \gamma^{\prime} C^{\mu\nu} + \delta^{\prime}
D^{\mu\nu}+\eta p^\mu p^\nu ~.
\end{equation}
Using $(\Delta^{-1})^{\mu \sigma} \Delta_\sigma\,^\nu= g^{\mu\nu}$ it
follows that the coefficient of $g^{\mu\nu}$ in $(\Delta^{-1})^{\mu
  \sigma} \Delta_\sigma\,^\nu$ should equal $1$ while the coefficients of
the other tensor structures, for example of $n^\mu n^\nu$, $n^\mu
p^\nu$ and $p^\mu p^\nu$, should vanish. Hence, we can determine the
coefficients in the propagator from the following equations
\begin{eqnarray}
\alpha^{\prime} &=& \frac{1}{p^2-\alpha}~, \\
(p^2-\alpha-\gamma)\gamma^{\prime}-\delta\,
\delta^{\prime}\frac{p^2({\bf p}^2 - (n\cdot
p)^2)}{\omega^2} &=& \frac{\gamma}{p^2-\alpha}~, \\
(p^2-\alpha-\gamma)\delta^{\prime} &=& \delta\,\beta^\prime
\frac{p^2}{\omega^2}~,\\
\frac{\delta}{p^2-\alpha}+\delta\,\gamma^\prime &=&
   (\omega^2-\beta)\delta^\prime\frac{p^2}{\omega^2}~, \\
\frac{1}{p^2}+\frac{\eta}{\lambda}p^2 &=& 0~.
\end{eqnarray}
Hence, we find that in covariant gauge the propagator in an
anisotropic plasma is given by
\begin{equation}
\Delta^{\mu\nu} = \frac{1}{p^2-\alpha} \left[A^{\mu\nu} - C^{\mu\nu}\right] +
\Delta_{G}\left[(p^2-\alpha-\gamma)\frac{\omega^4}{p^4}B^{\mu\nu} +
(\omega^2-\beta)C^{\mu\nu} + \delta\frac{\omega^2}{p^2}D^{\mu\nu}\right] -
\frac{\lambda}{p^4}p^\mu p^\nu ~,
\end{equation}
where
\begin{equation}
\Delta^{-1}_{G} = (p^2-\alpha-\gamma)(\omega^2-\beta) -
\delta^2 \left[{\bf{p}}^2-(n\cdot p)^2\right]~.
\end{equation}
For $\xi=0$, we recover the isotropic propagator in covariant gauge
\begin{equation}
\Delta^{\mu\nu}_{iso} = \frac{1}{p^2-\alpha}A^{\mu\nu} +
\frac{1}{(\omega^2-\beta)}\frac{\omega^4}{p^4}B^{\mu\nu} -
\frac{\lambda}{p^4}p^\mu p^\nu ~.
\end{equation}

\section{Heavy Quark Potential in an anisotropic plasma}
\label{mode-sec}

We determine the real part of the heavy-quark potential in the
nonrelativistic limit, at leading order, from
the Fourier transform of the static gluon propagator,
\begin{eqnarray}
V({\bf{r}},\xi) &=& -g^2 C_F\int \frac{d^3{\bf{p}}}{(2\pi)^3} \,
e^{i{\bf{p \cdot r}}}\Delta^{00}(\omega=0, \bf{p},\xi) \\
&=& -g^2 C_F\int \frac{d^3{\bf{p}}}{(2\pi)^3} \,
e^{i{\bf{p \cdot r}}} \frac{{\bf{p}}^2+m_\alpha^2+m_\gamma^2}
 {({\bf{p}}^2 + m_\alpha^2 +
     m_\gamma^2)({\bf{p}}^2+m_\beta^2)-m_\delta^4}~. \label{eq:FT_D00}
\end{eqnarray}
The masses are given by
\begin{eqnarray}
m_\alpha^2&=&-\frac{m_D^2}{2 p_\perp^2 \sqrt{\xi}}%
\left(p_z^2 {\rm{arctan}}{\sqrt{\xi}}-\frac{p_{z} {\bf{p}}^2}{\sqrt{{\bf{p}}^2+\xi p_\perp^2}}%
{\rm{arctan}}\frac{\sqrt{\xi} p_{z}}{\sqrt{{\bf{p}}^2+\xi p_\perp^2}}\right) \; , \\
m_\beta^2&=&m_{D}^2
\frac{(\sqrt{\xi}+(1+\xi){\rm{arctan}}{\sqrt{\xi}})({\bf{p}}^2+\xi p_\perp^2)+\xi p_z\left(%
p_z \sqrt{\xi} + \frac{{\bf{p}}^2(1+\xi)}{\sqrt{{\bf{p}}^2+\xi p_\perp^2}} %
{\rm{arctan}}\frac{\sqrt{\xi} p_{z}}{\sqrt{{\bf{p}}^2+\xi p_\perp^2}}\right)}{%
2  \sqrt{\xi} (1+\xi) ({\bf{p}}^2+ \xi p_\perp^2)} \; ,\\
m_\gamma^2&=&-\frac{m_D^2}{2}\left(\frac{{\bf{p}}^2}{\xi p_\perp ^2+{\bf{p}}^2}%
-\frac{1+\frac{2p_z^2}{p_\perp^2}}{\sqrt{\xi}}{\rm{arctan}}{\sqrt{\xi}}+\frac{
p_z{\bf{p}}^2(2{\bf{p}}^2+3\xi p_\perp^2)}{\sqrt{\xi}(\xi
p_\perp^2+{\bf{p}}^2)^{\frac{3}{2}}
p_\perp^2}{\rm{arctan}}\frac{\sqrt{\xi}
p_{z}}{\sqrt{{\bf{p}}^2+\xi p_\perp^2}}\right) \; ,\\
m_\delta^2&=&-\frac{\pi m_D^2\xi p_z p_\perp |{\bf{p}}|}{4(\xi
p_\perp^2+{\bf{p}}^2)^{\frac{3}{2}}}\, .
\end{eqnarray}
and
\begin{equation}
{\bf{p}}^2 = p_\perp^2 +p_z^2~.
\end{equation}
The above expressions apply when ${\bf n}=(0,0,1)$ points along the
$z$-axis; in the general case, $p_z$ and ${\bf p}_\perp$ get replaced by
$\bf{p\cdot n}$ and $\bf{p-n (p\cdot n)}$, respectively.

We first check some limiting cases. When $\xi=0$ then $m_\beta=m_D$
while all other mass scales in the static propagator vanish. Hence, we
recover the isotropic Debye potential
\beq
V({\bf{r}},\xi=0) = V_{\rm iso}(r) = -g^2 C_F\int \frac{d^3{\bf{p}}}{(2 \pi)^3}
\frac{e^{i{\bf{p\cdot r}}}}{{\bf{p}}^2+m_D^2} =
- \frac{g^2 C_F}{4 \pi r} \, e^{-\hat{r}}~,  \label{eq:V_iso}
\eeq
where $\hat{r}\equiv rm_D$.

Consider, on the other hand, the limit $r\to0$ for arbitrary
$\xi$. The phase factor in~(\ref{eq:FT_D00}) is essentially constant
up to momenta of order $|{\bf p}|\sim1/r$ and since the masses are
bounded as $|{\bf p}|\to \infty$ they can be neglected. The potential
then coincides with the vacuum Coulomb potential
\beq
V({\bf{r}}\to0,\xi) =
V_{\rm vac}(r) = -g^2 C_F\int \frac{d^3{\bf{p}}}{(2 \pi)^3}
\frac{e^{i{\bf{p\cdot r}}}}{{\bf{p}}^2}
= - \frac{g^2 C_F}{4 \pi r} ~.  \label{eq:V_vac}
\eeq
The same potential emerges for extreme anisotropy since all $m_i\to0$
as $\xi\to\infty$:
\begin{equation}
V({\bf{r}},\xi=\infty)= - \frac{g^2 C_F}{4 \pi r} ~.   \label{eq:V_xi=inf}
\end{equation}
This is due to the fact that at $\xi=\infty$ the phase
space density $f(\bf{p})$ from eq.~(\ref{eq:f_aniso}) has support only
in a two-dimensional plane orthogonal to the direction $\bf{n}$ of
anisotropy. As a consequence, the density of the medium vanishes in
this limit.

For an anisotropic distribution, the potential depends on the angle
between {\bf{r}} and {\bf{n}}. This can be seen analytically for small
but non-zero $\xi$. To linear order in $\xi$ the potential can be
expressed as
\begin{equation}
V({\bf{r}},\xi\ll1)= V_{iso}(r)-g^2 C_F \, \xi m_D^2 \int
\frac{d^3{\bf{p}}}{(2 \pi)^3} \, e^{i{\bf{p \cdot r}}} \,
\frac{\frac{2}{3}-({\bf p\cdot n})^2/{\bf{p}}^2}{({\bf{p}}^2+m_D^2)^2}~.
\end{equation}
For ${\bf{r}}$ parallel to the direction ${\bf{n}}$ of anisotropy,
\begin{eqnarray}
V({\bf{r}} \parallel {\bf{n}},\xi\ll1)
= V_{iso}(r)
\left[ 1+\xi
\left(2\frac{e^{\hat{r}}-1}{\hat{r}^2} -
\frac{2}{\hat{r}} -
1 -
\frac{\hat{r}}{6} \right) \right]~, \label{eq:V_r_parallel_n}
\end{eqnarray}
where $\hat{r}\equiv rm_D$, as before. This expression does not apply
for $\hat{r}$ much larger than 1, which is a shortcoming of the direct
Taylor expansion of $V({\bf r},\xi)$ in powers of $\xi$. However, for
$\hat{r}\simeq1$ the coefficient of $\xi$ is positive, $(\cdots) =
0.27$ for $\hat{r}=1$, and thus a slightly deeper potential than in an
isotropic plasma emerges at distance scales $r\sim1/m_D$.

When ${\bf{r}}$ is perpendicular to ${\bf{n}}$,
\begin{eqnarray}
V({\bf{r}}\perp{\bf{n}},\xi\ll1)= V_{iso}(r) \left[ 1 +  \xi
\left(\frac{1-e^{\hat{r}}}{\hat{r}^2 } +
\frac{1}{\hat{r}} +
\frac{1}{2} +
\frac{\hat{r}}{3}\right) \right]~.  \label{eq:V_r_perp_n}
\end{eqnarray}
The same limitations for $\hat{r}$ apply as in
eq.~(\ref{eq:V_r_parallel_n}). Here, too, the coefficient of the
anisotropy parameter is positive, $(\cdots) = 0.115$ for $\hat{r}=1$,
but smaller than for ${\bf{r}} \parallel {\bf{n}}$. Hence, a
quark-antiquark pair aligned along the direction of momentum
anisotropy and separated by a distance $r\sim 1/m_D$ is expected to
attract more strongly than a pair aligned in the transverse plane.

\begin{figure}
\includegraphics[width=7.4cm]{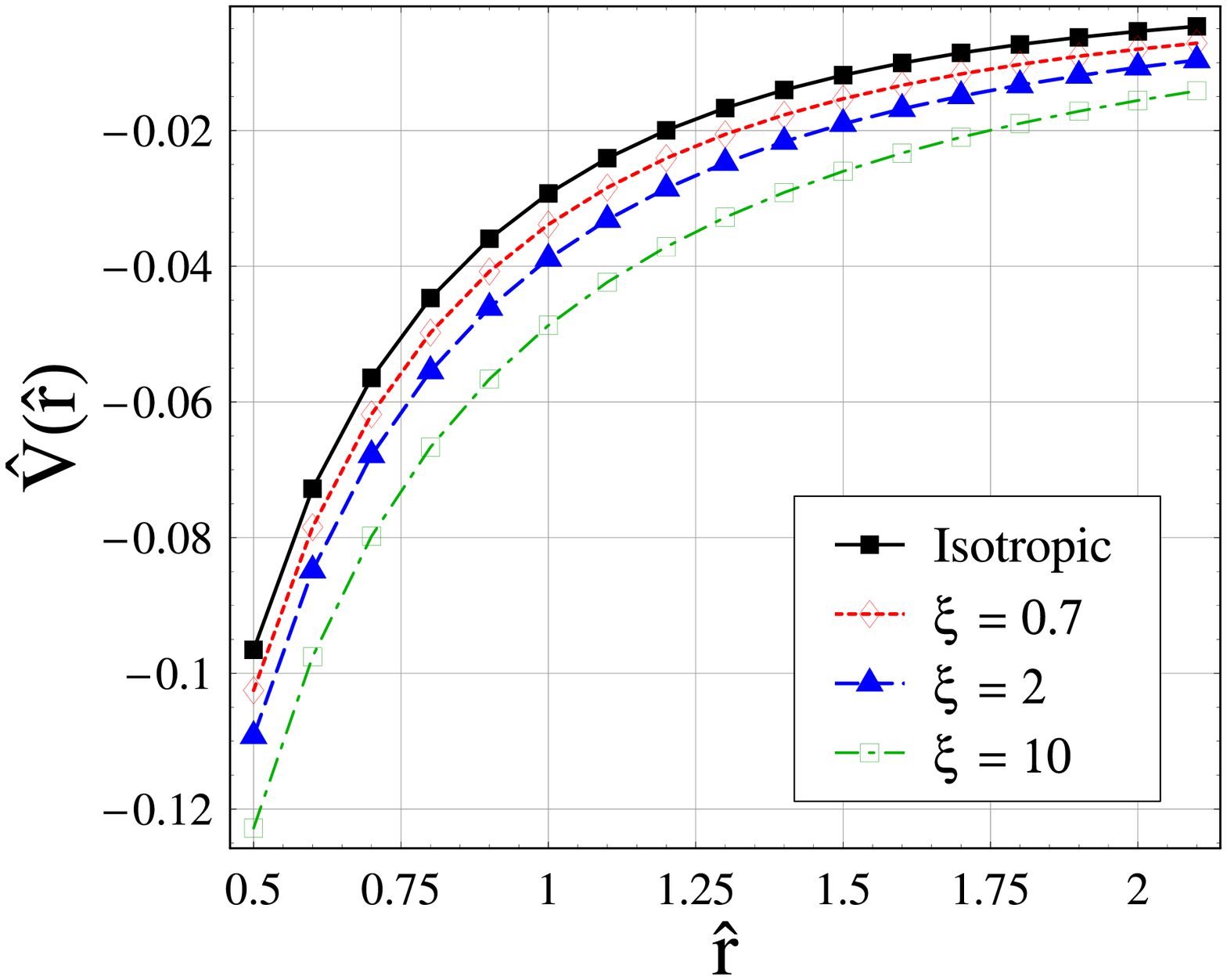}
\hspace{12mm}
\includegraphics[width=7cm]{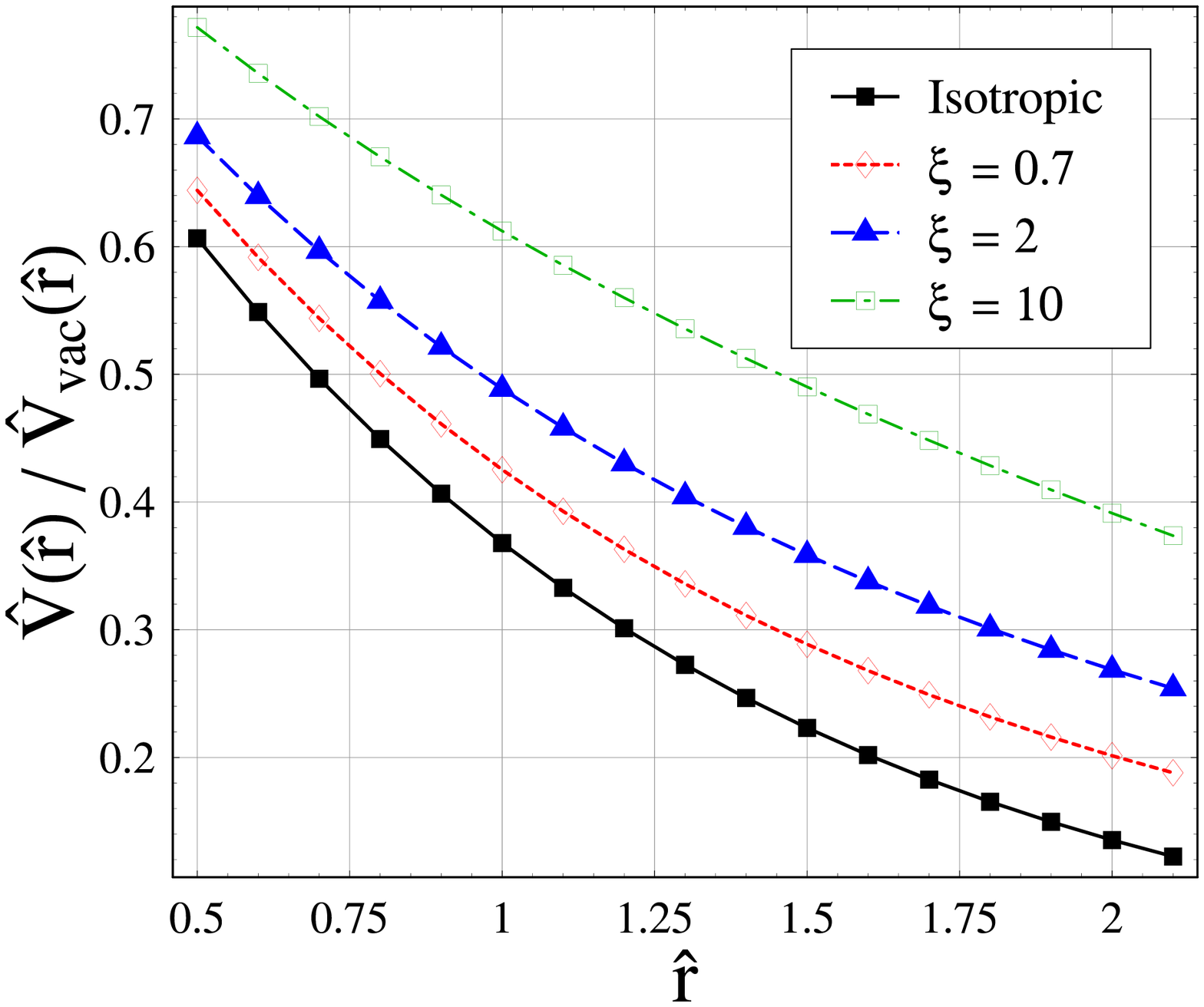}
\caption[a]{Heavy-quark potential at leading order as a function of
  distance ($\hat{r}\equiv rm_D$) for ${\bf{r}}$ parallel to the
  direction ${\bf{n}}$ of anisotropy. The anisotropy parameter of the
  plasma is denoted by $\xi$.\\ Left: the potential divided by the
  Debye mass and by the coupling, $\hat{V}\equiv V/(g^2 C_F m_D)$.
  Right: potential relative to that in vacuum.}
\label{fig:potential1}
\end{figure}

\begin{figure}
\includegraphics[width=7.4cm]{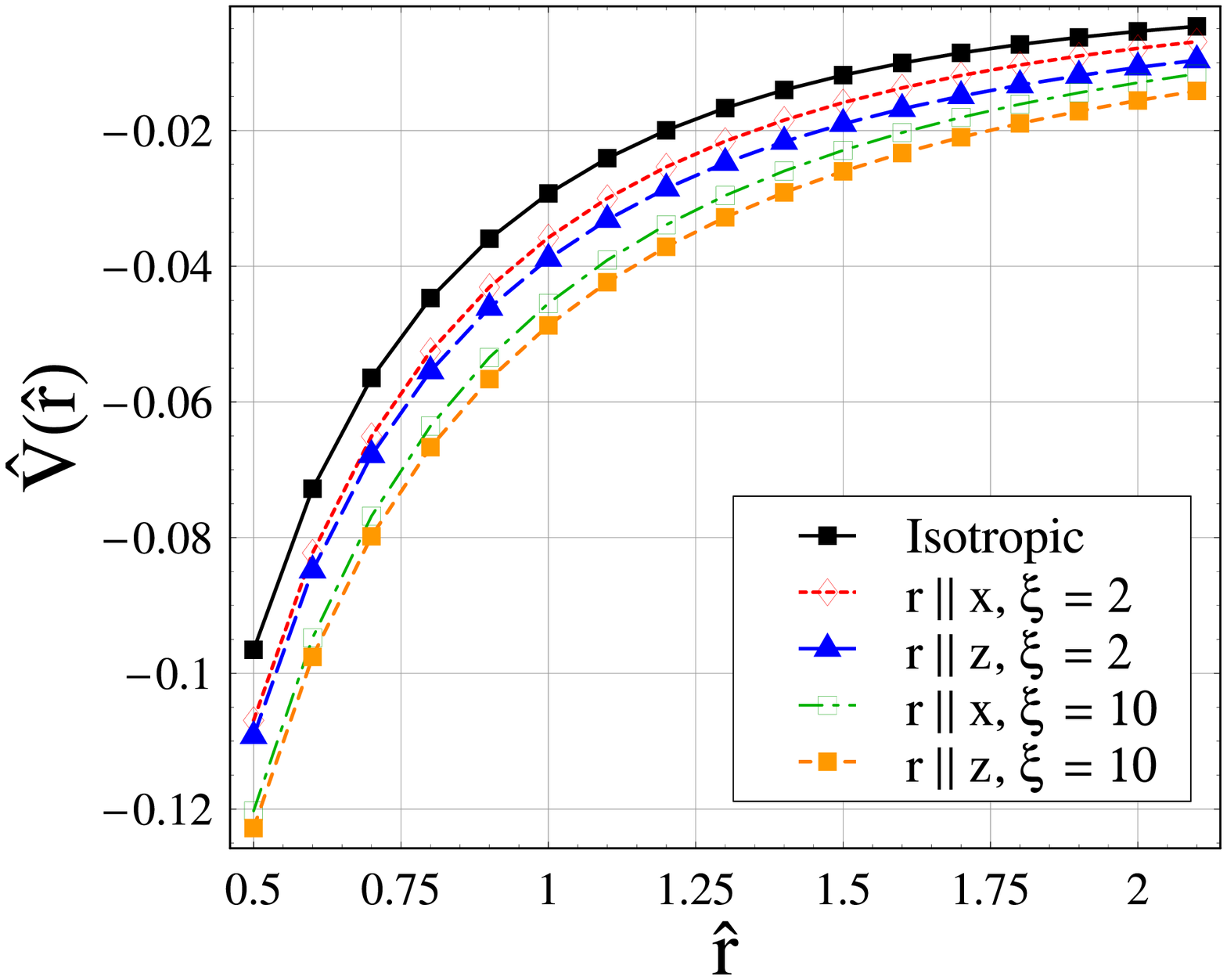}
\hspace{12mm}
\includegraphics[width=7cm]{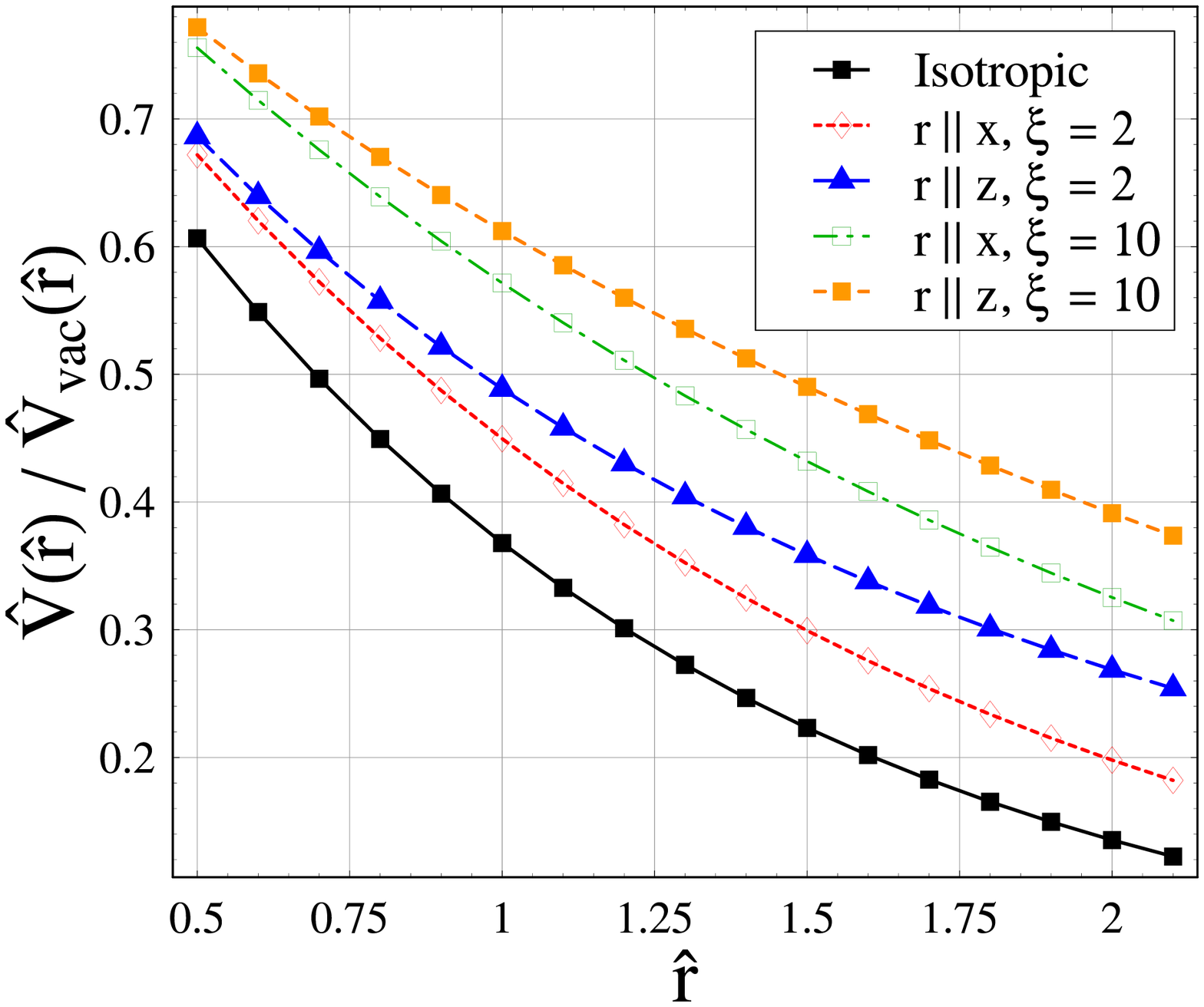}
\caption[a]{Comparison of $\hat{V}({\bf{r}} \parallel {\bf{n}},\xi)$
  and $\hat{V}({\bf{r}} \perp {\bf{n}},\xi)$.}
\label{fig:potential2}
\end{figure}

For general $\xi$ and $\hat{r}$, the integral in~(\ref{eq:FT_D00}) has
to be performed numerically. The poles of the function are
integrable\footnote{They are simple first-order poles which can be
  evaluated using a principal part prescription.}.
In Fig.~\ref{fig:potential1} we show the potential in the region
$\hat{r}\sim1$ for various degrees of plasma anisotropy. One observes
that in general screening is reduced, i.e.\ that the potential at
$\xi>0$ is deeper and closer to the vacuum potential than for an
isotropic medium. This is partly caused by the lower density of the
anisotropic plasma. However, the effect is not uniform in the polar
angle, as shown in Fig.~\ref{fig:potential2}: the angular dependence
disappears more rapidly at small $\hat{r}$, while at large $\hat{r}$
there is stronger binding for $\bf{r}$ parallel to the direction of
anisotropy. Overall, one may therefore expect that quarkonium states
whose wave-functions are sensitive to the regime $\hat{r}\sim1$ are
bound more strongly in an anisotropic medium, in particular if the
quark-antiquark pair is aligned along $\bf{n}$.


\section{Discussion and Outlook}
\label{conc-sec}

We have determined the HTL gluon propagator in an anisotropic
(viscous) plasma in covariant gauge. Its Fourier transform at
vanishing frequency defines a non-relativistic potential for static
sources. We find that, generically, screening is weaker than in
isotropic media and so the potential is closer to that in vacuum, in
particular if the $Q \overline{Q}$ pair is aligned along the direction
of anisotropy.

Our results are applicable when the momentum of the exchanged gluon is
on the order of the Debye mass $m_D$ or higher, i.e.\ for distances on
the order of $\lambda_D=1/m_D$ or less. For realistic values of the
coupling, $\alpha_s \approx 0.3$, $\lambda_D$ is approximately equal
to the scale $r_{\rm med}(T)\approx 0.5\, (T_c/T)$~fm introduced
in~\cite{Mocsy:2007yj,Kaczmarek:2004gv}, where medium-induced effects
appear.

Following the discussion in ref.~\cite{Mocsy:2007yj}, at short
distances, $r<r_{\rm med}(T)$, the potential is given by
\beq \label{eq:V_r<rmed}
V(r) \simeq -\frac{\alpha}{r} + \sigma r~,
\eeq
where $\sigma\simeq 1$~GeV/fm is the SU(3) string tension; color
factors have been absorbed into the couplings. Since $r_{\rm
med}(T)\sim 1/T$, it follows that at sufficiently high temperature
$r_{\rm med}(T)$ is smaller than $\sqrt {\alpha / \sigma}$ and so
the perturbative Coulomb contribution dominates over the linear
confining potential at the length scale $\lambda_D$.  Roughly, this
holds for $T\simge 2T_c$. In this case, our result is directly
relevant for quarkonium states with wavefunctions which are sensitive
to the length scale $\lambda_D\simeq r_{\rm med}$.

On the other hand, for lower $T$ the scale $r_{\rm med}(T)$ where
medium-induced effects appear may grow larger than $\simeq \sqrt
{\alpha / \sigma}$. In this regime, quarkonium states are either
unaffected by the medium; namely, if the quark mass is very large and
the typical momentum component in the wave function is $\gg 1/r_{\rm
  med}(T)$. Conversely, states with a root-mean square radius $\simge
r_{\rm med}(T)$ do experience medium modifications. For such states,
however, it is insufficient to consider only the (screened)
Coulomb-part of the potential which arises from one-gluon
exchange. Rather, one should then sum the medium-dependent
contributions due to one-gluon exchange {\em and} due to the
string~\cite{Mocsy:2007yj}. We postpone detailed numerical solutions
of the Schr\"odinger equation in our anisotropic potential to the
future. It will also be interesting to understand how the width of
quarkonium states~\cite{Mocsy:2007jz} which arises in HTL resummed
perturbation theory due to Landau damping of modes with low
frequency~\cite{Laine:2006ns} is affected by an anisotropy of the
medium.

\section*{Acknowledgments}
We acknowledge helpful discussions with A.~Mocsy and
P.~Petreczky. Y.G.\ thanks the Helmholtz foundation and the Otto Stern
School at Frankfurt university for their support. M.S.\ is supported
by DFG project GR 1536/6-1.


\end{document}